\documentclass[12pt,preprint]{aastex}
\received{2005}

\shorttitle{GCRT~J1745$-$3009}
\shortauthors{Hyman et al.}

\newcommand\src{\protect\objectname[]{GCRT~J1745$-$3009}}

\begin{document}
\title{A Faint, Steep Spectrum Burst from the Radio Transient \src}

\author{Scott D.~Hyman}
\affil{Department of Physics and Engineering, Sweet Briar College,
	Sweet Briar, VA  24595}
\email{shyman@sbc.edu}

\author{Subhashis Roy}
\affil{ASTRON, P.O. Box 2, 7990 AA Dwingeloo, The Netherlands.}
\email{roy@astron.nl}

\author{Sabyasachi Pal}
\affil{National Centre for Radio Astrophysics, Tata Institute of Fundamental Research, Pune, India}
\email{spal@ncra.tifr.res.in}

\author{T.~Joseph~W.~Lazio}
\affil{Remote Sensing Division, Naval Research Laboratory, Washington, DC
	20375-5351}
\email{Joseph.Lazio@nrl.navy.mil}

\author{Paul S.~Ray}
\affil{E.~O.~Hulburt Center for Space Research, Naval Research Laboratory, Washington, DC
	20375-5352}
\email{Paul.Ray@nrl.navy.mil}

\author{Namir E.~Kassim}
\affil{Remote Sensing Division, Naval Research Laboratory, Washington, DC
	20375-5351}
\email{Namir.Kassim@nrl.navy.mil}

\and

\author{Sanjay Bhatnagar}
\affil{National Radio Astronomy Observatory, Array Operations Center,
P.O. Box O,
1003 Lopezville Road,
Socorro, NM 87801-0387}
\email{bhatnagar@nrao.edu}

\begin{abstract}
\src\ is a transient bursting radio source located in the direction of the
Galactic center. It was discovered in a 330~MHz VLA observation from
2002 September 30--October 1 and subsequently rediscovered
in a 330~MHz GMRT observation from 2003 September 28 by Hyman et al. 
Here we report a new radio detection of the source in 330~MHz GMRT data taken on 
2004 March 20. The observed properties of the
single burst detected differ significantly from those measured previously, suggesting
that \src\ was detected in a new physical state. 
The 2004 flux density was $\sim$0.05 Jy,
$\sim$10$\times$ weaker than the single 2003 burst and $\sim$30$\times$ weaker than
the five bursts detected in 2002. 
We derive a very steep spectral index, $\alpha = -13.5 \pm 3.0$, 
across the bandpass, a new result previously not detectable due to limitations in the analysis of 
the 2002 and 2003 observations.
Also, the burst was detected for only $\sim$2 min.,
in contrast to the 10 min. duration observed in the earlier bursts.
Due to sparse sampling, only
the single burst was detected in 2004, as in the 2003 epoch, and we cannot rule out additional undetected bursts that may have occurred with the same
$\sim$77 min. periodicity observed in 2002 or with a different periodicity.
Considering our total time on source throughout both our archival
and active monitoring campaigns, we estimate the source exhibits detectable
bursting activity $\sim$7\% of the time.

\end{abstract}

\keywords{Galaxy: center --- radio continuum --- stars: variable: other}

\section{Introduction}\label{sec:intro}


\src\ is a novel transient radio source \citep{hlkrmy-z05,hlrrkn06}, whose
notable properties have included ``bursts'' with approximately 1~Jy peak flux density
lasting approximately 10~min.\ each and occurring at apparently
regular 77~min.\
intervals. We first identified \src\
from archival~330~MHz (90~cm) observations taken with the Very Large Array (VLA)
on~2002 September~30. We then detected a single burst from the source with the Giant
Metrewave Radio 
Telescope (GMRT) in archival observations from~2003 September~28 at the same frequency.
\src\ is located about~1.25\arcdeg\ south of the
Galactic center just outside (in angular distance) the shell-type
supernova remnant \objectname[]{SNR~G359.1$-$0.5} \citep{rf84}. The
environment of the source is discussed further in \cite{hlrrkn06}.

The source \src\ is notable because it is one of a small number of 
\emph{radio-selected} transients.  Moreover, with only a few
exceptions \citep{m02} such as electron cyclotron masers from flare
stars and the planets, plasma emission from solar radio flares, pulsar
radio emission, and molecular-line masers, most radio transients are
incoherent synchrotron emitters.  For an incoherent synchrotron
emitter, the energy density within the source is limited to an
effective brightness temperature of roughly $10^{12}$~K by the inverse
Compton catastrophe \citep{readhead94}.
The properties of \src\ suggest strongly that its brightness
temperature exceeds $10^{12}$~K by a large factor and that it is a
member of a new class of coherent emitters \citep{hlkrmy-z05}.

The two previous detections of \src\ were based on VLA and GMRT 330~MHz observations from
two different epochs, but from which similar source properties were observed.  
This \emph{Letter} reports on a third 330~MHz archival observation of \src, made with the 
GMRT in 2004, in which a single, much fainter and shorter burst is detected in contrast
to the burst properties observed in 2002 and 2003. In addition, the burst is found to have a very steep spectrum, as expected for a coherent emitter, providing another important clue to understanding the nature of this enigmatic source.

Possible models for \src\ include nearby objects such as a flaring brown
dwarf, flare star, or extra-solar planet, although, as discussed in
\cite{hlkrmy-z05}, the properties of the source do not easily fit these classes.
Alternative models proposed for \src\ thus far are a nulling pulsar (\cite{kp05}),
a double pulsar (\cite{tpt05}), a transient white dwarf pulsar 
(\cite{zhanggil}), and a precessing radio pulsar (\cite{zhuxu}). As noted in
the latter paper, the double pulsar model predicts a $\sim$3-yr bursting
activity period which is not supported by the redetections in 2003 and 2004.     
Furthermore, in \cite{hlrrkn06} we presented tentative evidence suggesting that the 2003 burst was
an isolated one, and not one of several bursts emitted at 77 min.
intervals, as observed in 2002.
This evidence could be problematic for some or all of 
these models. For example, the proposed double pulsar
would link the radio emission to a 77 min. orbital period. In this model, 
coherent radio emission is triggered by the shock formed through the interaction
of the wind of the more enigmatic pulsar with the magnetosphere of the companion pulsar.
On the other hand, in analogy to the PSR B1259-63 system which consists of a pulsar and Be star companion, 
it is also possible that the magnetosphere of the companion is not constant, and therefore that 
the radio bursts are not always triggered every orbit. 
Indeed, the much fainter detection from \src\ reported in this \emph{Letter} could  be
evidence of variable conditions in the environment around a companion star.

Similarly, the precessing radio pulsar and transient white dwarf pulsar models would also require one or more
types of nulling effects to explain the occurrence of isolated bursts in the short term,
as well as the low duty cycle in the long term. A few radio pulsars
are known to have a very large nulling fraction. PSR B1931+24 remains in an off state for $\sim$90\% of the time,
and it emits bursts quasi-periodically at $\sim$40 per day \citep{cordesetal04}.
Such a high nulling fraction may be consistent with the measured duty cycle estimated for \src.

\section{Observations and Results}\label{sec:observe}

The new, serendipitous detection reported in this paper is derived from 330
MHz GMRT Galactic center observations
obtained by two of us (S. Roy and S. Bhatnagar) as part of an unrelated
project and not included in \cite{hlrrkn06}. One of the observations,
from 2004 March 20-21, is pointed 0.5\arcdeg\ from \src\ and consists of
eleven 10 min. scans spread over six hours. 
The observations were carried out using the default observing mode with a
bandwidth of 16 MHz in each of the two available sidebands.  The sources
B1822-096 and B1714-25 were used as secondary calibrators. The GMRT does not
measure the system temperature ($T_{sys}$), and the increase in $T_{sys}$ from the
calibrator field to the target source affects the source visibility
amplitudes in the default observing mode (i.e., the Automatic Level
Control [ALC] in the system is turned on). We employed the following
method to correct for the $T_{sys}$ variation. As the system gain does not
change with the ALC off, we observed 3C48 and B1822-096 once with the ALC off and
determined the flux density of B1822-096 to be 10.8 Jy using the known flux density of
3C48 from \cite{baarsetal77}. Also with the ALC off, we determined the ratio of the
total power on the target source to that of B1822-096 to be 1.8.
Since this ratio was quite similar (within 10\%) for almost
all the antennas, rather than correcting the antenna-based gains, we
multiplied the final map of the source intensity distribution by this
value. We estimate the overall calibration uncertainty to be within
$\pm$15\%. The initial images were improved by phase and later amplitude
and phase self-calibration. To produce the final image, separately
self-calibrated data from both the sidebands were combined to improve
the uv-coverage.

We detect a $\sim$10$\sigma$ burst from \src\ in the middle of the first scan 
beginning at approximately 21:31:00 (IAT) on March 20 and lasting for approximately 2 min. The source
is unresolved and has a flux density of 57.9$\pm$6.6 mJy.
As a check on the calibration, we find the flux densities of several bright sources in the field-of-view
to be consistent with those obtained in the 330 MHz VLA survey of the Galactic center by \cite{nlkhlbd04}.

Over the past several years, we have conducted a 330 MHz search for radio transients in the Galactic
center \citep{hlkb02,hlknn03} using archival observations made between 1989 and the present,
and monitoring observations beginning in 2002.
\cite{hlrrkn06} summarize the VLA and GMRT observations and data reduction through 2005.
Most of these observations are pointed towards Sgr A* where the large primary beam at 330 MHz
(2.2\arcdeg [VLA] and 1.4\arcdeg [GMRT]) and high stellar density subtended optimize the
likelihood of
detecting a transient, including \src.
Our 2006 and 2007 monitoring program includes observations at 330 MHz (VLA) and simultaneous observations at 
235 and 610 MHz (GMRT); these will be described in a subsequent paper.

The flux density of the detected 2004 burst is much fainter than the 
$\sim$1~Jy peak flux density measured for
the five bursts detected in 2002 \citep{hlkrmy-z05} and for the one burst detected in 2003 \citep{hlrrkn06}.
Due to the relatively short and infrequent scans in the 2004
observations, we detect only a single burst, as in the 2003 epoch.
We cannot rule out the possibility that additional bursts were emitted at the same 77 min. interval 
observed in 2002. We note, however, that nondetections in additional scans made at multiples of 77 min. after the detected 2003 burst, but in which \src\ is located well beyond the half power point of the primary beam, suggest that the 2003 detected burst is actually an isolated one \citep{hlrrkn06}.

Figure~\ref{fig:lightcurve04} shows the light curves for all seven bursts  
detected to date.
While only the $\sim$2~min. decay portion of the 2003 burst was detected due to its coincidence with the beginning of a scan,
the 2004 burst is sampled completely within one scan. It is detected for 
only $\sim$2~min. compared with the $\sim$10~min. duration
observed for the bursts in 2002.


\src\ is unresolved in all three epochs. 
The resolution for the 2004 epoch is approximately 20\arcsec\ $\times$ 10\arcsec,
comparable to the 2003 epoch, and approximately a factor of~2
better than in the 2002 epoch. A Gaussian fit to the 2004
detection yields a position of
(J2000) right ascension $17^{\mathrm{h}}$ $45^{\mathrm{m}}$ 5\fs09 ($\pm
0\fs17$), declination $-30\arcdeg$ 09\arcmin\ 56.4\arcsec\ ($\pm 2\arcsec$),
which is consistent with the 2003 and 2002 positions and is approximately 
2.5$\times$ and 5$\times$ more accurate, respectively.
The source position and uncertainty cited above include
a correction for ionospheric refraction which is prevalent in low frequency observations and discussed in \cite{hlrrkn06} and \cite{nlkhlbd04}. 

\begin{figure}
\epsscale{0.7}
\rotatebox{-90}{\plotone{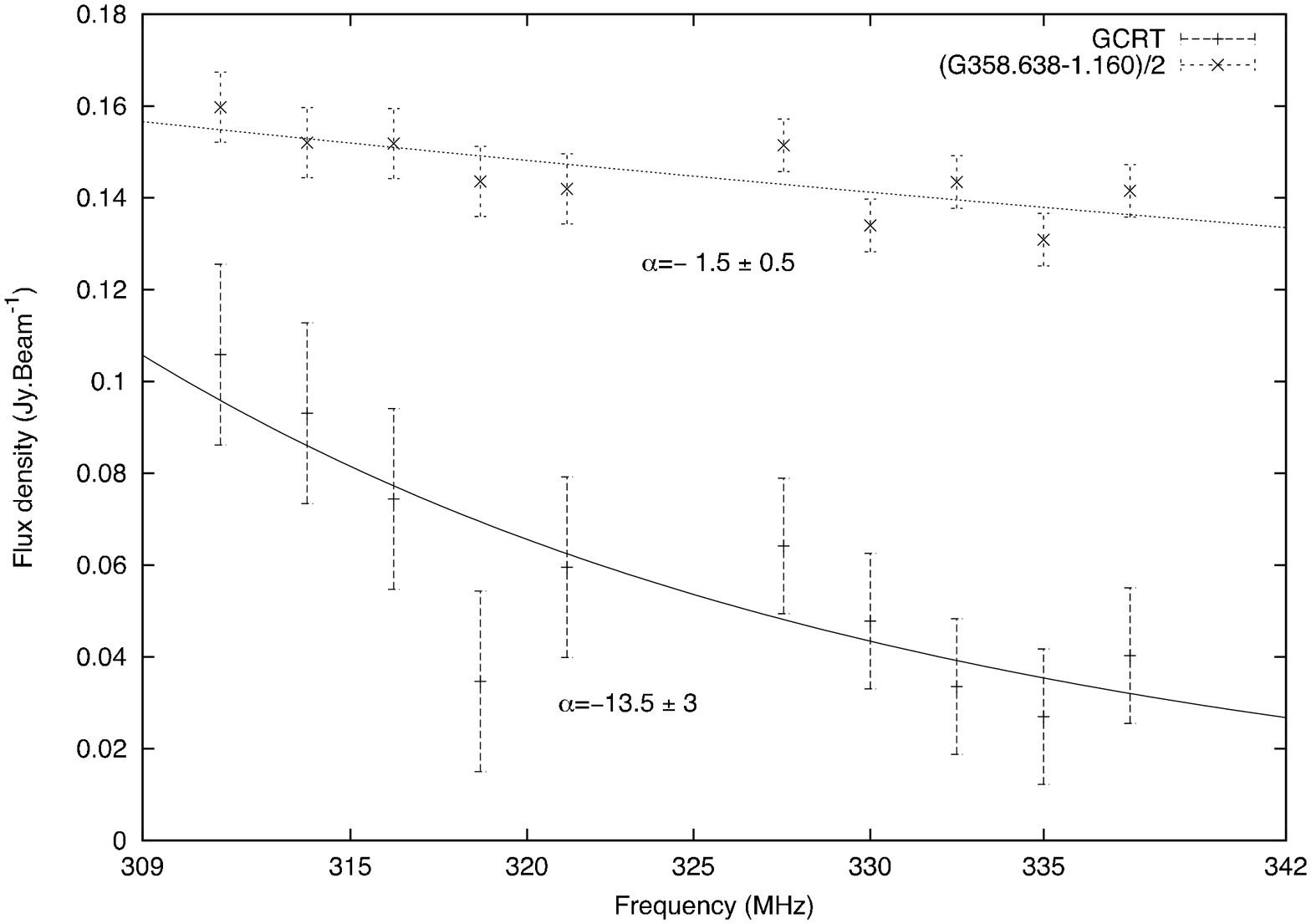}}
\caption{The spectra of \src\ and the nearby strong source,
G358.638$-$1.160. Pairs of adjacent channels were imaged across the 16 MHz 
bandpasses of the upper and lower sidebands. The derived spectrum for \src\
is $S \propto \nu^{-13.5 \pm 3.0}$. In comparison, the spectrum of
G358.638$-$1.160 is far less steep but is consistent with that obtained by \cite{nlkhlbd04}, indicating no significant errors in the bandpass calibration. Note that the flux densities plotted for G358.638-1.160 are shown scaled down by $\times$2.} 
\label{fig:spectrum04}
\end{figure}

Separate images were made for the upper (333 MHz) and lower (317 MHz) sidebands of the observations and yield flux densities of 42.1$\pm$7.2 mJy and 72.5$\pm$9.5 mJy, respectively. No significant differences are found in the shapes of the separate light curves generated for each sideband. Figure~\ref{fig:spectrum04} shows the spectrum of \src\ obtained by imaging pairs of adjacent frequency channels across the two sidebands. A power-law fit yields a very steep spectrum of $S \propto \nu^{-13.5 \pm 3.0}$ for \src. 
An identical analysis of the data for the nearby strong source G358.638$-$1.160 yields a spectral index of $-1.5 \pm 0.5$, consistent with the
determination of \cite{nlkhlbd04} who found a spectral index
of~$-1.2$ between~330 and~1400~MHz. 

A Monte-Carlo simulation was conducted to assess the confidence level of the steep spectrum obtained for the 2004 detection of \src. First, a spectrum was generated based on the fitted spectral index of $-13.5$ and normalized to the observed flux density at a particular channel. One thousand spectra were simulated by randomly adding Gaussian noise to the flux densities of the normalized spectrum, based on the observed noise level obtained for the individual channel images.
A flat spectrum fit to the \src\ data is ruled out with a confidence
level of 99.7\%.
Unfortunately, however, due
to limitations in the analysis of the 2002 and 2003 observations, we were not able to detect reliable evidence of a steep spectrum for those bursts.


No emission is detected from \src\ when imaging the 2004 observation at times when the burst is not occurring. 
From these 2004 observations, we are able to improve the
(5$\sigma$) upper limit for quiescent 330~MHz emission between
bursts to~6~mJy, as compared to earlier limits of~75~mJy and~25~mJy
from the 2002 and~2003 observations, respectively.
The upper limit on quiescent emission during
periods of no burst activity is 15~mJy at~330~MHz \citep{hlkrmy-z05}. In addition, we find an upper limit of 50\% circular polarization for the 2004 burst, while an upper limit of 15\% was previously determined for both the 2002 and 2003 bursts. 

\section{Discussion and Conclusions}\label{sec:Discussion}

In \cite{hlrrkn06} we crudely estimated the activity of the bursting
behavior of \src\ by comparing the time during which the source is
observed to be active to the total amount of observing time. With the addition of 
$\sim$40~hr of 330 MHz VLA observations analyzed in 2006, with no detection, the
total observing time is $\sim$120~hr.  The 2002 September~30 - October~1
bursts lasted for at least 6~hr, and since only a single burst was
detected on~2003 September~28 and on 2004 March 20, we assume that the source was active
for~$\sim$1~hr in each of these epochs. Thus, \src\ has exhibited bursting activity approximately 7\% of the
time it was observed.  

The 2002 and 2003 detections of \src\ occured 
in late September and the 2004 detection occurred in late March, suggesting
a roughly 6-month interval between active periods.
Although we find no detections in our 1998, 2002, 2005, and 2006 observations in March and September,
it is still possible that bursting activity did occur in these months but was either too 
faint or lasted too short a time to be detected. Similarly,
while our database has many observations in 2002, 2003, and 2006, we did not monitor for transients nearly
as often in 2004 and 2005, and useful archival data is very limited in those 
years. 

Since we have now detected \src\ in three significantly different observed states
($\sim$1~Jy versus $\sim$50~mJy burst strengths,
$\sim$10~min. versus $\sim$2~min. burst durations, and regularly repeating versus 
isolated bursts), future detections may yet exhibit additional properties
(e.g., emission in other frequency bands and polarized emission) that
will lead to a definitive understanding of the nature of \src. 
Given the high variability in flux
density already detected, and the very steep spectrum of the 2004 burst, it is even conceivable that bursts
much stronger than $\sim$1~Jy will be detected at 330 MHz and lower frequencies, although it is also possible
that the bursts are now continually decreasing in strength at all frequencies.

\begin{figure}
\epsscale{1.0}
\plotone{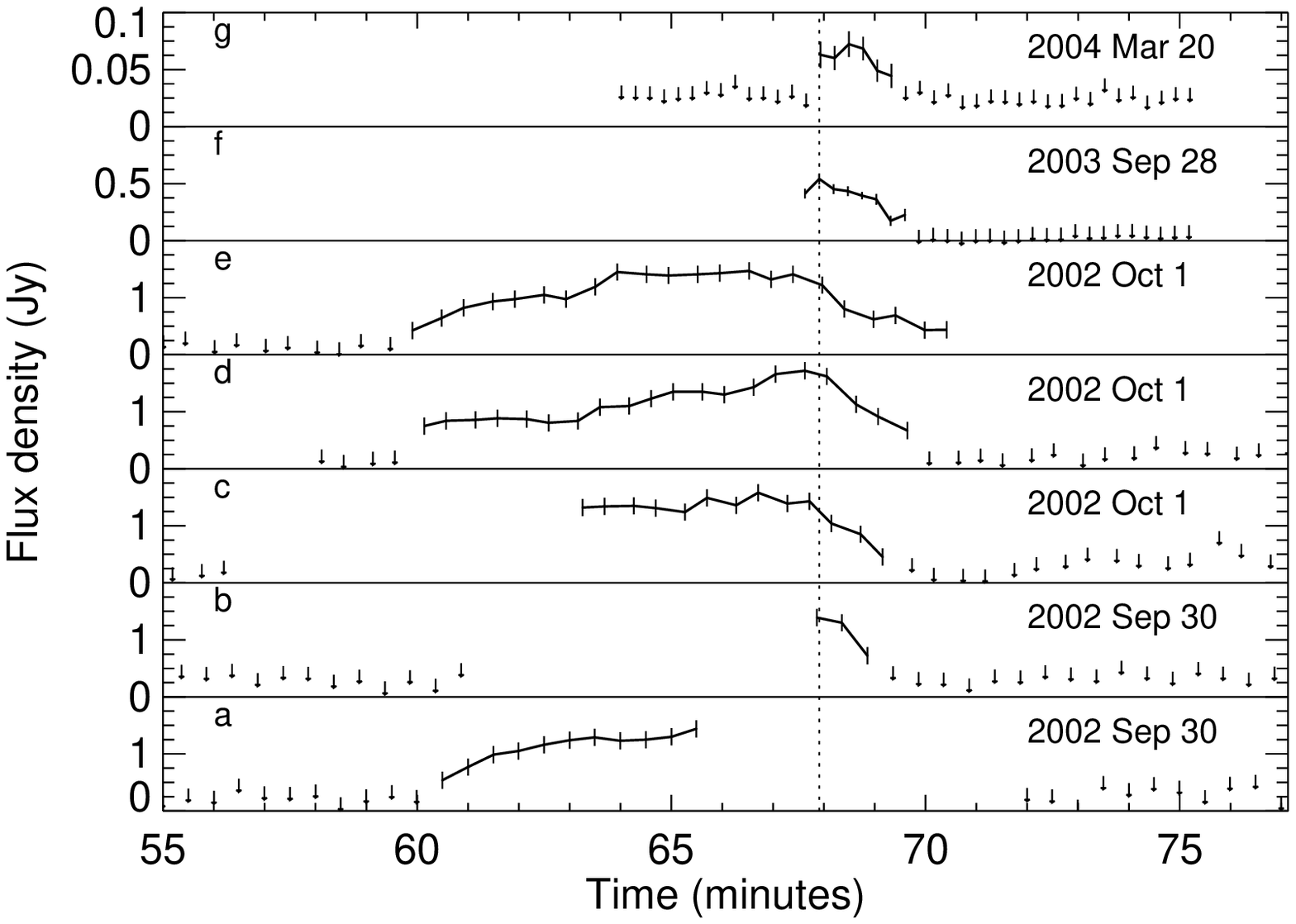}
\caption{The light curves of \src. The detected bursts from~2002 September~30
- October~1 are shown in panels a. - e. with 30-s
sampling \citep{hlkrmy-z05}. The single detected bursts from~2003
September~28 \citep{hlrrkn06} and~2004 March~20
(this \emph{Letter}) are shown in panels f. and g. with 17-s sampling
The y-axis ranges from 0 to 0.1~Jy and 0 to 1~Jy in the 2004 and 2003
panels, respectively, and from 0 to 2~Jy in the other
panels.
For the 2002 bursts, the light curve has been folded at the apparent
77.1~min.\ periodicity. For the 2004 and 2003 bursts,
the light curves have been
aligned in time to be consistent with the peak and decay portions of
the 2002
bursts. In many cases, because the existence of
\src\ was not known at the time of the observation, the full burst is 
not captured as the observations were interrupted for calibration or other targeted 
observations. The 2004 March~20 and 2003 September~28 observations consisted of one or two
$\sim$10~min.\ scans per hour for several hours. The arrows represent 3$\sigma$ upper limits
for nondetections. The vertical dotted line is placed at the fitted position
of the peak of the fourth 2002 burst for reference.} 
\label{fig:lightcurve04}
\end{figure}

We note that all three detections
suggest strongly that \src\ is a coherent emitter, as first indicated in \cite{hlkrmy-z05}, and now even more so by the very steep spectrum reported here.
Although the 2004 burst is much weaker than the previous bursts
it also appears to have much shorter rise and decay times, $\tau$, as seen in Figure~\ref{fig:lightcurve04}.
If we constrain the emitting region to be less than c$\tau$, with $\tau$ $\sim$1-min, then the brightness temperature of \src\ is
$\sim 10^{12}K (D/150 pc)^{2}$, where $D$ is the distance to the source. If the source is at the Galactic center, $\sim$8.5~kpc 
distant, its brightness temperature then still far exceeds the $10^{12}$K upper limit for an incoherent synchrotron emitter.





In summary, the new detection of \src\ consists of a single $\sim$50~mJy burst lasting only 2-min and exhibiting a very steep spectrum ($\propto \nu^{-13.5 \pm 3.0}$). The burst is 
significantly weaker and shorter than the $\sim$1~Jy and 10-min. bursts detected
in 2002 and 2003. Like the 2003 detection, the single burst detected in 2004
appears to be an isolated one, although the sparse sampling of the observation
does not rule out the possibility that additional bursts were emitted at the
same 77 min. period observed in 2002 \citep{hlkrmy-z05}.
Given the $\sim$120 hrs of 330 MHz observations searched, we estimate that the
source is in a detectable bursting state $\sim$7\% of the time.

Further, multi-wavelength observations are needed to better constrain the
physical nature of \src. Together with continued radio monitoring
for transient or weaker pulsed emission (e.g. pulsar-like), additional
desired observations include searches for quiescent or variable infrared or
X-ray counterparts.

\acknowledgements

The National
Radio Astronomy Observatory is a facility of the National Science
Foundation operated under cooperative agreement by Associated
Universities, Inc.  
We thank the staff of the GMRT that made these observations possible.
GMRT is run by the National Centre for Radio Astrophysics of the Tata
Institute of Fundamental Research.
S.D.H.\ is supported by funding from Research Corporation and SAO \emph{Chandra} grants GO6-7135F and GO6-7038B. Basic research in radio
astronomy at the Naval Research Laboratory is supported by 6.1 base funding.

\begin{thebibliography}{}



\bibitem[\protect\citeauthoryear{Baars et al.}{1977}]{baarsetal77}
	Baars, J.~W.~M., Genzel, R., Pauliny-Toth, I.~I.~K., \& Witzel, A.
	1977, \aap, 61, 99

\bibitem[\protect\citeauthoryear{Cordes et al.}{2004}]{cordesetal04} 
Cordes, J.~M., Lazio, T.~J.~W., McLaughlin, M.~A. 2004, New \ Astron. \ Rev., 48, 1459

\bibitem[\protect\citeauthoryear{Frail \& Kulkarni}{1991}]{fk91}
	Frail, D.~A.\ \& Kulkarni, S.~R.  1991, \nat, 352, 785

\bibitem[\protect\citeauthoryear{Gaensler et al.}{2004}]{gaensleretal04}
	Gaensler, B.~M., van der Swaluw, E., Camilo, F., Kaspi, V.~M.,
	Baganoff, F.~K., Yusef-Zadeh, F., \& Manchester, R.~N.  2004, \apj,
	616, 383

\bibitem[\protect\citeauthoryear{Hyman et al.}{2006}]{hlrrkn06}
	Hyman, S.~D., Lazio, T.~J.~W., Roy, S., Ray, P.~S., Kassim, N.~E.,
	Neureuther, J.~L.  2006, \apj, 639, 348

\bibitem[\protect\citeauthoryear{Hyman et al.}{2005}]{hlkrmy-z05}
	Hyman, S.~D., Lazio, T.~J.~W., Kassim, N.~E., Ray, P.~S.,
	Markwardt, C.~B., \& Yusef-Zadeh, F.  2005, \nat, 434, 50

\bibitem[\protect\citeauthoryear{Hyman et al.}{2003}]{hlknn03}
	Hyman, S.~D., Lazio, T.~J.~W., Kassim, N.~E., Nord, M.~E., \&
	Neureuther, J.~L.  2003, 
	{Astron. Nachr.}, 324, 79

\bibitem[\protect\citeauthoryear{Hyman et al.}{2002}]{hlkb02}
	Hyman, S.~D., Lazio, T.~J.~W., Kassim, N.~E., \& Bartleson,
	A.~L.  2002, 
	\aj, 123, 1497

\bibitem[\protect\citeauthoryear{Kulkarni \& Phinney}{2005}]{kp05}
	Kulkarni, S.~R. \& Phinney, E.~S. 2005, \nat, 434, 28



\bibitem[\protect\citeauthoryear{Melrose}{2002}]{m02}
	Melrose, D.~B.  2002, 
	Publ.\ Astron.\ Soc.\ Aust., 19, 34

\bibitem[\protect\citeauthoryear{Nord et al.}{2004}]{nlkhlbd04}
	Nord, M.~E., Lazio, T.~J.~W., Kassim, N.~E., Hyman, S.~D.,
	LaRosa, T.~N., Brogan, C.~L., \& Duric, N.  2004, 
	\aj, 128, 1646

\bibitem[\protect\citeauthoryear{Readhead}{1994}]{readhead94}
	Readhead, A.~C.  1994, \apj, 426, 51

\bibitem[\protect\citeauthoryear{Reich \& F\"urst}{1984}]{rf84}
	Reich, W.\ \& F\"urst, E.  1984, \aaps, 57, 165

\bibitem[\protect\citeauthoryear{Turolla, Possenti, \& Treves}{Turolla
	et al.}{2005}]{tpt05}
	Turolla, R., Possenti, A., \& Treves, A.  2005, \apj, 628, L49
	press; astro-ph/0506199


\bibitem[\protect\citeauthoryear{Zhang \& Gil}{2005}]{zhanggil}
	Zhang, B. \& Gil, J. 2005, \apj, 631, L143

\bibitem[\protect\citeauthoryear{Zhu \& Xu}{2005}]{zhuxu}
	Zhu, W.~W. \& Xu, R.~X. 2005, \mnras, 365, L16


\end{thebibliography}
\end{document}